\documentclass{iau}

\usepackage{amsmath}
\usepackage{graphicx}
\usepackage{multirow}
\usepackage{natbib}
\usepackage{xspace}

\providecommand\gaia{\textit{Gaia}\xspace}
\providecommand\gdr[1]{\gaia~DR#1\xspace}
\providecommand\edr[1]{\gaia~EDR#1\xspace}
\providecommand\grvs{\ensuremath{G_\mathrm{RVS}}\xspace}
\providecommand\gbp{\ensuremath{G_\mathrm{BP}}\xspace}
\providecommand\grp{\ensuremath{G_\mathrm{RP}}\xspace}
\providecommand\bpminrp{\ensuremath{(\gbp-\grp)}\xspace}
\providecommand\vect[1]{\ensuremath{\mathbf{#1}}\xspace}
\providecommand\muas{\ensuremath{\mu\text{as}}\xspace}
\providecommand\msun{\ensuremath{M_\odot}\xspace}
\providecommand\mjup{\ensuremath{M_\mathrm{Jup}}\xspace}
\providecommand\muaspyr{\ensuremath{\mu\text{as yr}^{-1}}\xspace}
\providecommand\kms{\ensuremath{\text{km~s}^{-1}}}
\providecommand\selfun{\ensuremath{S}}

\begin{document}

\lefttitle{Anthony G.A.~Brown}
\righttitle{The \gaia mission}

\jnlPage{1}{7}
\jnlDoiYr{2025}
\doival{10.1017/xxxxx}
\volno{395}
\pubYr{2025}
\journaltitle{Stellar populations in the Milky Way and beyond}

\aopheadtitle{Proceedings of the IAU Symposium}
\editors{J. Mel\'endez,  C. Chiappini, R. Schiavon \& M. Trevisan, eds.}

\title{\gaia: Ten Years of Surveying the Milky Way and Beyond}

\author{Anthony G.A.~Brown}
\affiliation{Leiden Observatory, Leiden University, Einsteinweg 55, 2333 CC Leiden, the Netherlands}

\begin{abstract}
On January 15 2025, the \gaia mission completed the collection of the astrometric, photometric, and spectroscopic data
for about $2.5$ billion celestial sources, from the solar system to the Milky Way to the distant universe. Work is
ongoing to produce \gdr{4} based on the first $5.5$ years of data, with the release expected in 2026. The full $10.5$
year survey will be turned into \gdr{5} which will open up scientific possibilities beyond \gdr{4}. In this contribution
I give a brief overview of the \gaia mission, summarize results from the GaiaUnlimited project, provide a glimpse of
what is to come in \gdr{4}, and summarize the new science opportunities that \gdr{5} will bring. I close with a look
ahead at the successor to \gaia, the GaiaNIR mission, which will survey the Milky Way in the infrared, thus probing the
Galactic ecosystem in the regions hidden to the \gaia mission.
\end{abstract}


\maketitle

\section{Overview of the \gaia mission}

\gaia was launched on 19 December 2013 and started its survey of the skies on July 25 2014 after the spacecraft and
payload commissioning period \citep{2016A&A...595A...1G}. The end of \gaia's science observations was on January 15
2025, completing a $10.5$ year survey of the solar system, the Milky Way, and beyond. \gaia collected accurate
positions, parallaxes and proper motions for all sources to magnitude $G\sim21$, its white-light photometric band
covering the range $330$--$1050$~nm \citep{2021A&A...649A...2L}. Next to the $G$-band photometry, multi-colour
photometry was obtained for all sources through two low-resolution fused-silica prisms dispersing all the light entering
the field of view \citep{2021A&A...649A...3R, 2023A&A...674A...2D, 2021A&A...652A..86C}. The Blue Photometer (BP)
operated in the wavelength range $330$--$680$ nm, while the Red Photometer (RP) covered the wavelength range
$640$--$1050$ nm.  The photometry is needed to account for colour dependent effects in the astrometry and is used to
estimate astrophysical parameters for all sources \citep{2023A&A...674A..26C}. \gaia also carried a medium resolution
($R\sim11\,000$) spectrograph on board, the Radial Velocity Spectrometer (RVS), which covered the wavelength range
$847$--$874$ nm \citep{2018A&A...616A...5C}.  The RVS spectra are used to determine radial velocities for sources to
$G\sim17$ \citep{2023A&A...674A...5K} and to extract astrophysical parameters, including detailed abundance information.
for the brighter stars at $\grvs\lesssim14$ \citep{2023A&A...674A..29R}.

Starting in 2016, the Gaia Collaboration regularly published data releases based on increasing observation time spans. It
should be stressed here that the data releases are publicly available world-wide without any proprietary period for the
teams in charge of the data processing. The most recent releases were \gdr{3} in 2022 \citep{2023A&A...674A...1G} and the
Focused Product Release in 2023. In addition to the core scientific data products mentioned above, the releases feature
a large set of value-added data products that turn the \gaia catalogues into a true supermarket for astronomers.
Examples are the classification and characterization of variable stars \citep{2023A&A...674A..13E}; the astrophysical
characterization of \gaia sources \citep[see][and references therein]{2023A&A...674A..26C} the large catalogue of
non-single stars \citep{2023A&A...674A..34G}; the astrometry and orbits of solar system objects
\citep{2023A&A...674A..12T}; the extragalactic contents of the \gaia catalogue \citep{2023A&A...674A..31D}; and numerous
specific data products, such as astrometry from a special observing mode for crowded fields \citep{2023A&A...680A..35G},
gravitationally lensed quasars \citep{2024A&A...685A.130G}, diffuse interstellar bands \citep{2023A&A...680A..38G}, 
radial velocity time series for long period variables \citep{2023A&A...680A..36G}, and orbits for solar system objects
based on $5.5$ years of \gaia astrometry \citep{2023A&A...680A..37G}.

For a detailed overview of the \gaia mission refer to \cite{2016A&A...595A...1G}. For an overview of \gaia astrometry in a
brief historical context and an introduction to the astrometric data processing and the potential sources of systematic
errors see \cite{2021ARA&A..59...59B}, which also summarizes a wide range of scientific results from \gdr{2} (as of
2020).  \cite{2020ARA&A..58..205H} provides a complementary review of the insights on the Milky Way's early history
obtained from \gaia's first two data releases. A number of more recent reviews relevant to the topic of stellar
populations include \citet[][galactic archaeology]{Deason_2024}, \citet[][stellar streams]{Bonaca2025}, \citet[][open
clusters]{Cantat_2024}, \citet[][binaries]{ElBadry_2024}, and \citet[][white dwarfs]{Tremblay_2024}.

\section{The GaiaUnlimited project}

\begin{figure}[t]
    \includegraphics[width=\textwidth]{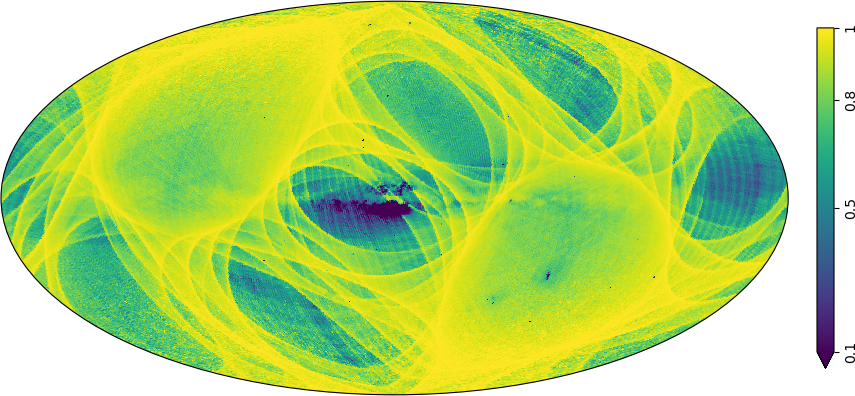}
    \caption{The completeness of the \gaia survey at $G=20.9$ as a function of position on the sky. The map shows
        completeness as the probability of a source being included in the \gaia catalogue given its celestial position.
        The map uses an Aitoff projection in galactic coordinates with the Galactic centre in the middle and
        $\ell=+180^\circ$ on the left. The map is at HEALPix level 7 \citep{2005ApJ...622..759G} with pixels of
    $0.2$~deg$^2$.} \label{fig:m10compl}
\end{figure}

Like any survey of the sky, the \gaia survey is limited in completeness and subject to a number of selection effects.  A
very basic illustration of this are the maps of the large scale three-dimensional distribution of Milky Way stars that
have been constructed based on stellar distances estimated from \gaia parallaxes combined with photometric information
\cite[eg.][]{BailerJones2021,Anders2022}. The maps show an artificial concentration of stars around the sun and a clear
limitation in the distance over which the Milky Way disk and halo spatial structure can be probed directly.  Although in
principle the \gaia survey is magnitude limited only, there are numerous survey details that complicate the selection
function, including the survey strategy (scanning law), data losses due to events on the spacecraft, crowding of sources
on the sky, the complications in linking individual detections to sources on the sky, the many decisions on which data
to include in the processing for a given \gaia data product, and finally, any subsample selection or data quality
constraints imposed on the \gaia catalogue by the user of the data. This results in a selection probability dependent on
many source properties, including position on the sky, apparent brightness in $G$, colour \bpminrp, astrophysical
parameters, etc.  The selection function gives the probability that a given source on the sky will be present in the
\gaia catalogue.  Figure \ref{fig:m10compl} shows the \gdr{3} selection probability for sources at $G=20.9$ as a
function of of position on the sky \citep{Cantat2023}. The patterns due to the \gaia scanning law are visible as well as
the effects of source crowding near the Galactic centre and in the direction of the Magellanic clouds and globular
clusters.

The first work to consider the \gaia selection function in detail was by \cite{Bovy2017} who presented an inventory of
the stellar content of the solar neighbourhood based on \gdr{1}. After \gdr{2} appeared, a major effort was led by
D.~Boubert and A.~Everall to develop a statistical model of the \gaia survey selection function which accounts for the
scanning law effects and data losses. They presented selection functions and corresponding tools to use these for
\gdr{2} \citep{Boubert2020} and \edr{3} \citep{Everall2022}. A completeness assessment focused on the \gdr{2} RR Lyrae
catalogue \citep{Clementini2019,Rimoldini2019} was presented by \cite{Mateu2020}. These selection functions were
based on a `reverse engineering' of the \gaia catalogue or on comparison to other survey data as there was no access to
information on the details of data losses or data processing decisions internal to the \gaia Data Processing and Analysis
Consortium (DPAC). It was an oversight at the time of the creation of the DPAC that no task was included to derive the
selection function. This state of affairs was discussed at the \gaia Sprint workshop at the KITP in Santa Barbara in
2019 and led to the creation of the GaiaUnlimited project\footnote{\url{https://gaia-unlimited.org/}} which is aimed at
determining the \gaia survey selection function and providing corresponding data and tools to the community. The project
is embedded in DPAC which makes it possible to use additional non-public information of relevance to the construction of
the selection function.

The GaiaUnlimited approach to deriving the \gaia selection function is laid out in \cite{Rix2021}. Their figure 1
schematically shows the steps from observation and data collection in a given survey to the production of a catalogue
from which a sample of sources may be extracted by someone for their scientific analysis. The survey consists of
observations of sources according to some strategy and with a detection efficiency that depends on the
telescope/instrument combination, position on the sky, wavelength range, etc. This is followed by data processing to
turn the raw data into astrophysical information, in the case of \gaia a set of attributes \vect{q} for each source. The
list of sources with their attributes is the so-called `parent catalogue'. Ideally one traces all the steps from source
detection to the derivation of \vect{q}, accounting for any data losses due to, e.g., events on the spacecraft or on
ground that interrupt the observations, or decisions taken during data processing on which raw or intermediate data to
include. This would then produce the selection function $\selfun^\text{parent}(\vect{q})$ of the parent catalogue.  In
practice this approach is not feasible, so \cite{Rix2021} propose to derive $\selfun^\text{parent}(\vect{q})$ either
through a statistical modelling of the selection function based on a high-level understanding of the survey and data
processing \citep[the approach taken by][]{Boubert2020, Everall2022}, or through the use of deeper or higher-resolution
surveys which can be considered as `ground truth' against which to calibrate models of
$\selfun^\text{parent}(\vect{q})$.

An important aspect of the selection function is that it should provide selection probabilities for both actual objects
and for arbitrary, or counterfactual, objects that may have attributes \vect{q} not present in the catalogue itself (for
example objects beyond the faint limit of \gaia). This allows to incorporate the selection function in forward modelling
of the data, for example when using Bayesian inference methods. Ideally the selection function should depend on `simple
observables' such as sky position, apparent brightness, colour, and parallax, and any model used in the data analysis
should make predictions of these quantities (which are derived from astrophysical attributes such as distance, effective
temperature, chemical composition, surface gravity). An important choice made in \cite{Rix2021} is that the effects of
dust on source observability are considered part of the modelling of the attributes \vect{q} (i.e.\ the model used to
analyze the data should predict the correct magnitudes and colours for sources affected by dust along the line of
sight).

The selection function tools developed by GaiaUnlimited are accessible through
GitHub\footnote{\url{https://github.com/gaia-unlimited/gaiaunlimited}} and are described in several publications.
\cite{Cantat2023} present an empirical model for the \gdr{3} parent survey selection function which depends only on one
parameter per patch of sky, the median magnitude $M_{10}$ for which sources in the catalogue have less than 10
astrometric transits ($\texttt{astrometric\_matched\_transits}\leq10$). This parameter is computable from the \gaia
catalogue and closely tracks the completeness limit and the steepness of the completeness function at the faint end of
the survey. The model was calibrated using the Dark Energy Camera Plane Survey \citep[DECaPS,][]{shlafly2018,
saydjari2023} and validated with \textit{Hubble} Space Telescope observations of globular clusters. Using the \gdr{3}
parent survey selection function as a starting point it is straightforward to derive selection functions for subsets of
the \gaia catalogue, as described in \cite{CastroGinard2023}.

A set of papers from GaiaUnlimited present more complex cases of constructing and using selection functions. In all
cases the paper is accompanied by code and notebooks to reproduce the results. \cite{Cantat2024} construct the
selection function for a sample of Red Clump stars that was selected from the combination of the APOGEE DR17
\citep{majewski2017, abdurrouf2022} and \gdr{3} GSP-Spec \citep{2023A&A...674A..29R} data sets. The sample has dense
coverage of the disk near the sun combined with the longer reach of the APOGEE pencil beams. This joint sample was used
to study the density profile of mono-abundance populations in different metallicity ranges, showing that the density
profiles steepen with metallicity and that there is metallicity-dependent flaring of the $[\alpha/\text{Fe}]<0.1$ disk.
\cite{CastroGinard2024} took the popular application of using the \gaia catalogue RUWE parameter to select potential
binaries and derived a sky-position-dependent threshold and showed how selection on RUWE translates to a selection on
the underlying physical properties (orbital parameters, distances) of the binary population. Tools are provided to
repeat the selection function study for any simulated binary population. \cite{Khanna2024} presented a sample of Red
Clump stars selected from the \gdr{3} and AllWISE \citep{wright2010,mainzer2011} catalogues. The joint sample was used
to study the overall structure of the Milky Way's disk. A flared thin disk and short thick disk without flaring provide
a good description of the spatial distribution of the Red Clump sample. The model residuals show clear signatures of the
warp of the disk and, interestingly, also of the spiral arms. Finally \cite{Kurbatov2024} constructed a sample of metal
poor RGB stars ($[\text{M}/\text{H}]<-1.3$) selected from the \cite{Andrae2023} sample, which consists of \gdr{3} stars
for which data driven metallicities were derived from the combination of BP/RP spectra and CatWISE
\citep{2021ApJS..253....8M}, using APOGEE stellar parameters as the training data. \cite{Kurbatov2024} studied the
spatial distribution of the Aurora population, one of the oldest components of the Milky Way, formed in situ
\citep{Belokurov2022}. They modeled the Aurora population with a parametrized spatial distribution, a globular cluster
RGB luminosity function, and included the effects of extinction and distance smearing due to parallax uncertainties. In
combination with the \gaia selection function for this sample they derived the parameters of the spatial distribution of
Aurora. The paper thus presents a fairly complex demonstration of the use of selection functions in data analysis, with
the corresponding tools available online.

Work on the statistical modelling of the selection function is ongoing (Castro-Ginard et al., in prep.) and will be used
for the production of the \gdr{4} selection function.

\section{Upcoming \gaia data releases}

The publication of \gdr{4} is planned no earlier than the middle of 2026 and in early 2025 the status of the data
processing for this release is that the astrometric, photometric, and spectroscopic (radial velocity) data processing
have finished. This forms the input for the production of the more advanced data products such as: the classification and
characterization of variable sources; astrophysical characterization of all sources; the processing of solar system
object data; the analysis of non-single star data (binaries, exoplanets); and the characterization of extragalactic
sources (galaxy surface brightness profiles, QSO host galaxy properties, and the detection of lensed QSO systems). The
lengthy data schedule for the data release is driven by the need to keep systematic errors in all data products well
below the precision levels, which increase as more and more data is included in the processing. This involves a very
significant effort on improving calibrations, developing new approaches for newly identified systematics, and the
validation and documentation of the results.

The astrometric results show the expected improvements from \gdr{3} to DR4 (34 vs.\ 66 months of data), factors of
$\sqrt{2}$ and $2\sqrt{2}$ improvement across all magnitudes for the parallaxes and proper motions, respectively. At the
bright end ($G<13$) the parallax uncertainties show even larger improvements, which reflects the better control over
systematics. The formal parallax uncertainties vary from $\sim10$~\muas at the bright end ($6<G<13$) to $\sim400$~\muas
at $G=20$. For the proper motions the corresponding numbers are $\sim5$~\muaspyr and $\sim200$~\muaspyr. In addition to
the standard astrometric catalogue, astrometry will also be provided for nine crowded sky regions using the approach
described in \cite{2023A&A...680A..35G}. These regions are in Baade's Window and Sagittarius I in the Bulge, the Large
and Small Magellanic Clouds, and in the globular clusters Omega Centauri, 47 Tuc, M22, NGC 4372, M4 \citep[see table 1
in][]{2023A&A...680A..35G}.

The integrated photometry (fluxes in the broad bands $G$, \gbp, \grp) shows improvements, in particular at the bright
end ($G<13$) where the uncertainties are closer to the limit expected in the absence of calibration errors \citep[cf.\
Fig.\ 14 in][]{2021A&A...649A...3R}, reflecting again the improved control over systematics. The BP/RP spectra have also
further improved in quality and this includes the calibration of the photometric passbands. These improvements will
translate to improved astrophysical parameters inferred from the photometry and BP/RP spectra.

The data processing for the RVS instrument now extends to the faintest magnitudes, $14\lesssim\grvs\lesssim16.5$. At
these brightness levels spectra from the individual $4.4$ second exposures are very noisy, but stacking the spectra
observed over the 66 months time baseline of \gdr{4} allows to bring out the Ca triplet lines and still determine radial
velocities even at the faintest end. The precision of the radial velocities will also improve by the above mentioned
factor $\sqrt{2}$ due to the doubling of the number of observations. A much improved identification of spurious radial
velocities will also be provided, making kinematic studies with \gdr{4} more robust.

\gdr{4} will feature two major changes with respect to the previous releases, the publication of all data processing
results for all sources in the internal source list and the publication of all epoch (time series) data. For previous
\gaia data releases quality limits were set on which sources would enter the various tables in the published catalogue,
for example the main \gdr{3} source table is populated with sources that are brighter than $G\sim21$, have a sufficient
number of astrometric observations, and have astrometric uncertainties below a certain limit \citep[see][section
4.4]{2021A&A...649A...2L}. For \gdr{4} no such filtering on quality will be done and the data processing results for all
$\sim2.5$ billion sources will be published.  However, a clear distinction will be made between the roughly 2 billion
sources considered most reliable (corresponding to what would have been published in previous releases) and the
remaining sources for which the results may be less reliable. Detailed guidance will be provided so that users of the
\gdr{4} data can make informed choices about the samples of sources they may wish to use for their scientific
applications.

The other major change is the publication of all the epoch data, this means time series of astrometric, photometric, and
radial velocity measurements, and of BP/RP and RVS spectra for all sources in \gdr{4}. In previous releases time series
data was only provided for photometry \citep[$\sim12$ million sources in \gdr{3},][]{2023A&A...674A..13E} and for a very
limited subset of sources also radial velocity time series were provided \citep[e.g.][]{2023A&A...680A..36G}. The
availability of all epoch data thus represents a vast increase in data volume and in the richness of science
applications that will be possible. For example the epoch astrometry can be used in searches for exoplanets or stars
orbiting massive objects such as black holes and neutron stars. A prominent example is the discovery of \gaia BH3
\citep{2024A&A...686L...2G} for which the epoch astrometric data was provided along with the publication of the discovery.
Indeed \gdr{4} will already contain the results of the non-single star processing, featuring a large list of exoplanet
candidates, ordinary binary stars, and systems with compact objects. Nevertheless there will be plenty of scope for the
community to make their own discoveries among the vast number of astrometric time series. The same holds for the other
time series data, where combinations of different types of data will increase the discovery potential. A taste of what
is possible with epoch BP/RP spectra was provided in a \gaia Image of the
Week\footnote{\url{https://www.cosmos.esa.int/web/gaia/iow_20181115}}. There is a lot to look forward to in \gdr{4}!

\begin{figure}[t]
    \includegraphics[width=0.48\textwidth]{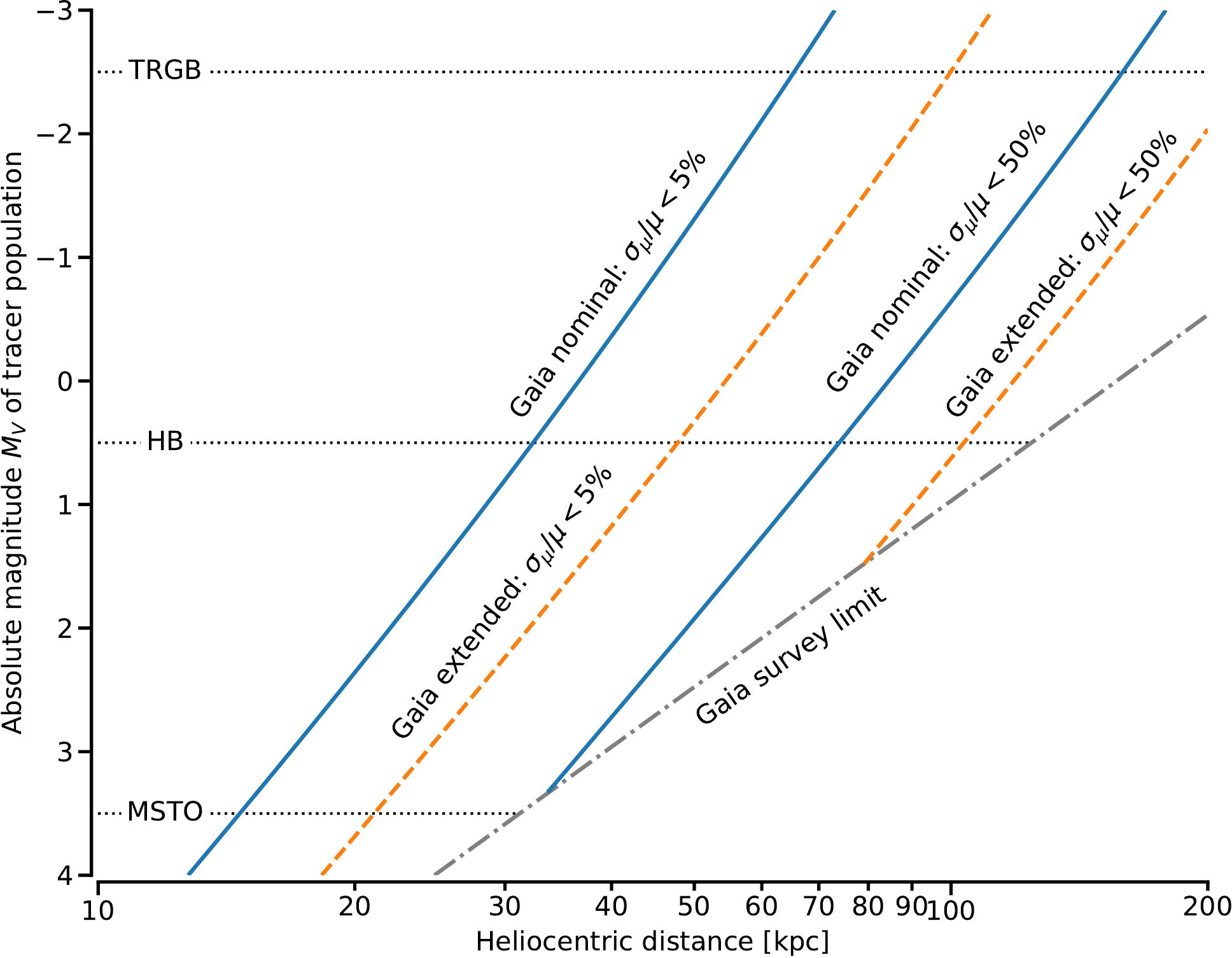}\hfil
    \includegraphics[width=0.48\textwidth]{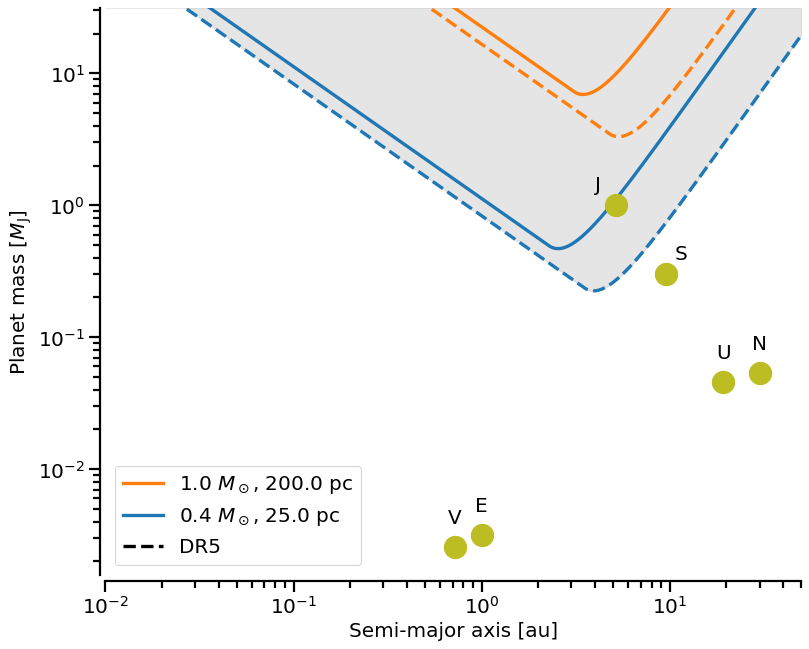}
    \caption{Left: Predicted relative proper motion uncertainties as a function of distance for stars with the absolute
        magnitudes along the vertical axis for \gdr{4} (solid blue lines) and \gdr{5} (dashed orange lines). Three
        specific tracer populations are indicated: the main sequence turnoff, horizontal branch and tip of the red giant
    branch stars. After figure 9 in \cite{2021ARA&A..59...59B}. Right: \gaia exoplanet discovery space (gray coloured
    area) assuming a $50$~\muas single epoch precision for the astrometry and a detection threshold at a signal-to-noise
    of 20. The solid lines show the \gdr{4} detection limits in the mass vs.\ semi-major axis space for a $1$~\msun and
    $0.4$~\msun host star, the dashed lines are for \gdr{5}. The green dots indicate planets in the solar system.}
    \label{fig:dr5science}
\end{figure}

\gdr{5} will be based on the data collected over the full extended \gaia mission, doubling the observational time
baseline from the nominal $5$ years to $10.5$ years. This opens up a range of new science cases that cannot be addressed
with \gdr{4}. The basic mission products, positions, parallaxes, photometry, and radial velocities increase in precision
by a factor $t^{0.5}$ (i.e.\ $\sim40$\% improvement in precision), while proper motions increase much more rapidly in
precision with $t^{1.5}$, a factor of about $2.8$ from \gdr{4} to \gdr{5}. This translates into a volume increase over
which proper motions at a certain precision are accessible by a factor $\sim23$. Notable science cases that can be
addressed with \gdr{5} include: a massive improvement in the orbits of solar system object, with one or more orbital
periods covered out to the Trojans and the Hildas; increase by a factor of 20 of the volume over which internal
kinematics of open clusters can be studied to the $0.3$--$0.5$~\kms accuracy level in tangential velocity; upper limits
to the energy flux density of gravitational waves in the $<1$--$2$ nHz frequency may be obtained; and a much improved
measurement of the acceleration of the solar system barycentre \citep{ssbaccel}. Two specific cases are further
illustrated in Fig.~\ref{fig:dr5science}. The left panel shows the distances out to which proper motions have relative
accuracies of 5 and 50\%. This large volume increase enabled by \gdr{5} allows studying the internal kinematics of bright
populations in dwarf galaxies out to $\sim100$~kpc and precision tangential motion studies of the dynamically unmixed
halo beyond 20~kpc. The right panel of Fig.~\ref{fig:dr5science} illustrates the \gaia exoplanet discovery space in the
planetary mass vs.\ orbital semi-major axis plane. \gaia provides an astrometric exoplanet survey of about
$10^6$--$10^7$ stars, unbiased across spectral type, age, and chemical composition of the primary. For \gdr{4} thousands
of exoplanet discoveries are predicted in the $<15$~\mjup range around A to M stars. Figure \ref{fig:dr5science} shows
that \gaia will discover long period Jupiter mass planets, providing excellent synergies with the PLATO mission
\citep{2014ExA....38..249R, 2017AN....338..644M} and opportunities for ground-based imaging follow-up. The doubling of
the time baseline with \gdr{5} provides access to giant exoplanets on orbital periods comparable to the giant planets in
the solar system. The number of discovered exoplanets could triple with \gdr{5}. In summary, \gdr{5} will not just be an
improved version of \gdr{4} but will open up a whole range of new science opportunities.

\section{Beyond the \gaia mission}

The first proposal for the \gaia mission was submitted to the European Space Agency (ESA) in 1993, four years ahead of
the appearance of the Hipparcos catalogue in 1997. This illustrates how the thinking about future improvements to
observational facilities is often triggered even before the facilities have come to full fruition. \gaia is no
exception, and the first thoughts on future all-sky astrometric survey missions were developed already before the launch
of \gaia as part of the response to an ESA call for science themes for its L2 and L3
missions\footnote{\url{https://sci.esa.int/s/wNRzBdw}}. The science case for all-sky nano-arcsecond level astrometry was
laid out in \cite{2013AstrometryL2L3} together with a discussion on the challenges involved \citep[see section 6.4
in][for a summary]{2021ARA&A..59...59B}. At the meeting where the L2/L3 proposals were presented to ESA's advisory
bodies, the proposal for a second \gaia or an infrared version of \gaia was also presented. The concept of an infrared
version of \gaia, GaiaNIR, was subsequently taken up by the community in Europe as the most practical option to pursue
and a first full design study was conducted in 2016. This led to the submission of a white paper to ESA's Voyage 2050
call for science themes for the long term planning of ESA's science programme \citep{2021ExA....51..783H}. The GaiaNIR
science case was selected as one of four options for large-class science missions for the time frame 2035--2050.

GaiaNIR aims to build on the success of the \gaia mission by extending the survey concept into the near-infrared (NIR)
wavelength range which will allow to study the dust-obscured regions of the Milky Way, in particular in the disk. An
important lesson from \gaia is that the dense phase space sampling complemented by astrophysical information is
essential to uncover subtle features of the Milky Way's dynamics, history, and stellar populations, which for example
makes possible the study of `galactoseismology' \citep{2021ARA&A..59...59B}. In particular, densely sampling the regions
of the disk where star formation takes place would allow to directly study the link between Milky Way dynamics and star
formation.  GaiaNIR's probing of the stellar populations in the Milky Way disk all the way to the Bulge, bar and central
regions, will provide a much more complete view of how the dynamics in the Milky Way's inner regions influences regions
further out in the disk through resonances and radial migration. Closer to home, GaiaNIR would enable an astrometric
census of planets around ultra-cool dwarfs close to the Sun and the combination of GaiaNIR and \gaia would provide access
to exoplanets with orbital periods in the $\sim30$--$35$ year range. The proper motions derived from the \gaia-GaiaNIR
combination will reach sub-\muaspyr accuracy levels, corresponding to tangential velocities of a less than 1~\kms at
100~kpc, thus opening up a whole new precision space in the halo of the Galaxy. The combination of the two missions
will also serve to improve the precision of the celestial reference frame and maintain it over much longer periods. This
is crucial for many future applications of astrometry across a wide range of astronomical topics, and is essential to
accurate pointing of large telescopes on ground and in space. See \cite{2021ExA....51..783H} for an overview of many
more science cases that can be addressed with GaiaNIR, including also the solar system and extragalactic science. The
gains to be made by probing the dust-obscured regions of the Milky Way are nicely illustrated by two recent studies, the
study by \cite{2024A&A...692A.194G} uncovering many star clusters by an analysis of the GLIMPSE survey
\citep{2003PASP..115..953B} and the study by \cite{2025A&A...693A..28A} mapping Mira variables on the far side of the
Galactic disk by using the VVV \citep{minniti2010}, VVVx \citep{minniti2018}, and WISE \citep{wright2010} surveys.

A detailed design of the GaiaNIR mission was developed during a study conducted by ESA with participation from
the scientific community and made public as a
report\footnote{\url{https://sci.esa.int/web/future-missions-department/-/60028-cdf-study-report-gaianir}}. The study
started from the principles of the design of \gaia and Hipparcos, thus a revolving scanning survey concept employing two
telescopes separated by a large basic angle \citep[cf.][section 3]{2021ARA&A..59...59B}. The main challenge at the time
in using the $RJHK$ bands wavelength range ($\sim800$--$2500$~nm) was the lack of time-delayed-integration (TDI) capable
IR detectors. The capability to operate in TDI mode is essential to the scanning concept. The design study was focused
on a mission that would fit in an M-class mission cost envelope which led to dropping the spectrophotometric (BP/RP) and
spectroscopic (RVS) instruments from the design. With the then foreseen mirror sizes and detector options, the
performance of GaiaNIR was similar to that of \gaia albeit with the advantage of observing many more objects by being
able to penetrate the dust in the Milky Way. Since then several developments took place that make GaiaNIR an even more
exciting prospect than it already was.

IR detector technology has progressed and there is now the option to use avalanche photodiodes (APDs) operating in the
NIR which offer TDI capabilities and a much better overall noise performance. These are a real game-changer for GaiaNIR,
the TDI capability leading to a much simplified mission design, and the lower noise allows to reach better performance
than \gaia even for similar size mirrors, enabling a deeper and more accurate survey. The science themes addressed by
GaiaNIR are recommend as a large mission class option in ESA's Voyage 2050 which opens up the possibility to introduce
spectrophotometric as well as a spectroscopic capabilities in the GaiaNIR design. This would offer \gaia BP/RP like low
resolution spectra for the $\sim10$ billion sources predicted to be observed by GaiaNIR as well as radial velocities and
abundance information for hundreds of millions of stars, something out of reach for even the largest future ground-based
spectroscopic surveys. The enormous success of \gaia rests on the vast amount of astrophysical information that
complements its exquisite astrometry, and this will be essential also to the success of GaiaNIR. A final improvement of
GaiaNIR would be to consider a larger ($\sim3$~m) mirror size for the telescopes which can be achieved by using
segmented mirror technology and would mitigate the increased susceptibility to crowding due to the longer operating
wavelengths of GaiaNIR.

A paper in preparation by Hobbs et al.\ will detail the latest design of GaiaNIR, including the developments mentioned
above. Members of the stellar population community are very welcome to get involved in further developing the science
case and design of this mission. More information and contact details can be found on the GaiaNIR
website\footnote{\url{https://www.astro.lu.se/GaiaNIR}}.

Between now and a possible launch of GaiaNIR there are plenty of opportunities for further advances in Milky Way
astrometry, in particular by exploiting the combination of \gaia and other surveys from ground or in space. The
combination of \gdr{3} and Euclid \citep{2024arXiv240513491E} already provides two epochs of high precision astrometry
over a time baseline of $\sim7$--$8$ years which can lead to an improvement over the \gdr{3} proper motion precisions by
a factor 10, as illustrated by the study of \cite{libralato2024} who derived proper motions from the \gaia-Euclid
combination for the globular cluster NGC 6397, reaching precisions better than $100$~\muaspyr at $G=19$--$20$. In
addition they can derive proper motions for faint \gaia stars for which only a position is available in \gdr{3}.
Likewise the Roman mission will produce precise astrometry \citep{2019JATIS...5d4005W} which can be combined with \gaia
to derive much more precise proper motions. Both Euclid and Roman will provide multi-epoch astrometry which allows the
derivation of proper motions for a vast number of sources beyond the reach of \gaia. In the future this data can be
combined with GaiaNIR to further improve proper motions over long time baselines. 

Many other opportunities for synergies between \gaia, GaiaNIR and other ground and space-based surveys are described in
\cite{2021ARA&A..59...59B} and \cite{2021ExA....51..783H}.

\begin{figure}[t]
    \includegraphics[width=\textwidth]{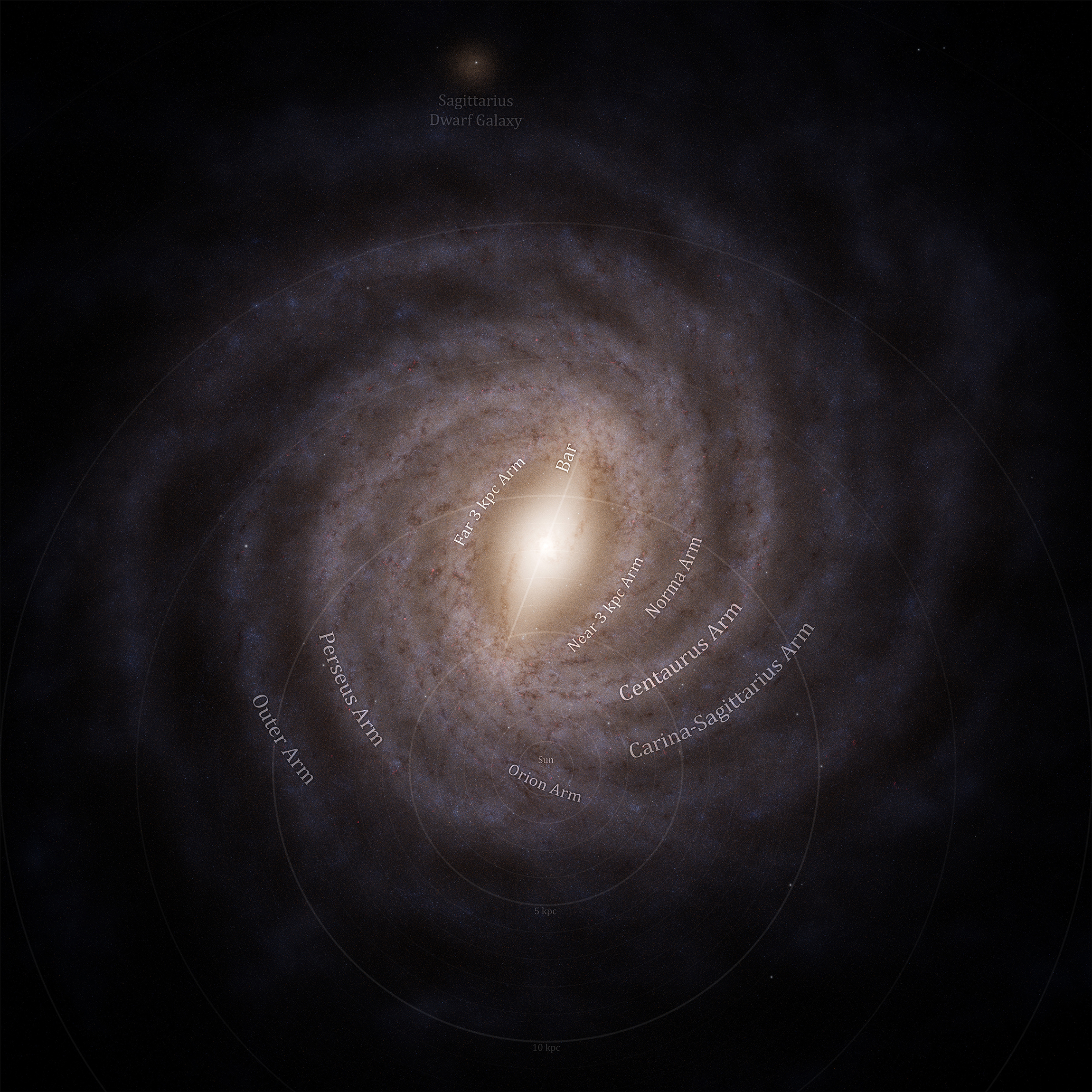}
    \caption{Artist's impression of the Milky Way based on the latest knowledge derived from the \gaia catalogue data.
    Credits: ESA/Gaia/DPAC, Stefan Payne-Wardenaar, CC BY-SA 3.0 IGO. Download different versions of the image here:
\url{https://www.esa.int/ESA_Multimedia/Images/2025/01/The_best_Milky_Way_map_by_Gaia}.}
    \label{fig:mw}
\end{figure}

\section{End of \gaia science observations}

The astrometric precision of \gaia rests to a large extent on the exquisite control over the spin rate and attitude of
the spacecraft. The spin rate is synchronized with the motion of sources across the focal plane, such that that the TDI
observing mode produces sharp images. The attitude control serves to smoothly execute the \gaia scanning law which
allows for precise attitude modelling as part of the astrometric solution. The spin-rate and fine attitude control are
ultimately managed through a micro-propulsion system consisting of micro-Newton-level thrusters. The fuel for this
micro-propulsion system was originally budgeted for a $6.5$ year mission lifetime, including half a year for
commissioning and a possible one-year mission extension. Over the course of the first years of spacecraft operations it
became clear that there was enough fuel onboard to keep the micro-propulsion system running for potentially up to
$10$--$11$ years.  This knowledge was used to argue for a five year extension to the nominal \gaia mission, to reach a
total time baseline of $10+$ years, allowing for the additional \gdr{5} release and the unlocking of new science
opportunities. 

The fuel for the micro-propulsion system was predicted to be exhausted by early 2025 and it was decided to end \gaia
science observations on January 15 2025 at 06:15 UTC. The little bit of fuel remaining for the micro-propulsion system
was used for technology tests that were intended to better understand the \gaia S/C and payload, in particular with the
goal to get a better understanding of the basic angle variations, which can potentially lead to improved control over
the systematic errors in the astrometry caused by these variations \citep[cf.][]{2021A&A...649A...2L,
2021ARA&A..59...59B}. For some of the technology tests \gaia was put into different orientations with respect to the
Sun, which made the spacecraft appear much brighter in the sky (down to magnitude $\sim15.5$) and as part of a `farewell
campaign' the astrophotography community worldwide was encouraged to take pictures of \gaia and submit these to a
website for
display\footnote{\url{https://www.cosmos.esa.int/web/gaia/ground-based-observations-of-gaia-spacecraft-2025}}. The
occasion of the end of \gaia science observations was also an opportunity to promote a beautiful new artist's impression
(see Fig.~\ref{fig:mw}) of the Milky Way based on the latest knowledge of our galaxy obtained from the \gaia data.

On March 27 2025 \gaia was passivated and put onto an orbit around the sun, away from its long time home at the second
Lagrange point. Although this marked the formal end of the operational phase of the \gaia spacecraft, the \gaia mission
continues onward to \gdr{4} and \gdr{5}!

\bibliographystyle{iaulike}
\bibliography{brown}

\end{document}